\documentclass[12pt,a4paper]{article} 
\usepackage[dvipsnames,usenames]{color}
\usepackage{multirow}
\usepackage{ftnxtra}
\usepackage{subeqn}
\usepackage[numbers,square,comma,sort&compress]{natbib}
\usepackage{lscape}
\usepackage{graphicx}

\begin{document}

\title{The general static spherical perfect fluid solution with \\
EoS parameter $w=-1/5$}

\author{\.{I}brahim 
Semiz\thanks{mail: ibrahim.semiz@boun.edu.tr} \\ \\
Bo\u{g}azi\c{c}i University, Department of Physics\\
34342 Bebek, \.{I}stanbul, TURKEY}
    
\date{ }

\maketitle

\begin{abstract}
The general analytical solution for the static spherically symmetric metric supported by a perfect fluid with isothermal (proportional) equation-of-state $p = w \rho$ is not known at the time of this writing, except for the trivial cases $w=0$ and $w=-1$; and for $w=-1/3$. We show that if Buchdahl coordinates are used, the problem becomes analytically solvable for $w=-1/5$; display and discuss the solution(s), and exhibit the connection of this case to the $w=-1$ case.
\end{abstract}


\section{Introduction: SSSPF solutions with isothermal EoS}
\label{sect:Intro}

Contrary to the impression in a non-negligible fraction in the literature (see works citing~\cite{Kiselev}), the solution for ``blackhole surrounded by quintessence'' is not known~\cite{Visser_on_Kiselev,Lake_on_Kiselev,Semiz_on_Kiselev}. To be more precise, the {\it general} exact static spherically symmetric (SSS) solution of Einstein's Field Equations (EFE)
\begin{equation}
G_{\mu\nu} = \kappa T_{\mu\nu}               \label{ee}
\end{equation}
are unknown for perfect fluid (PF) source, i.e.
\begin{equation}
T_{\mu\nu} = (\rho + p) u_{\mu} u_{\nu} + p g_{\mu\nu}  \label{pfemt}
\end{equation}
 where $\rho$ and $p$ satisfy
\begin{equation}
p=w\rho.     \label{eos}
\end{equation}
with constant $w$.
In eq.(\ref{ee}), $G_{\mu\nu}$ is the Einstein tensor, for whose definition we use conventions of~\cite{mtw}; $T_{\mu\nu}$ the stress-energy-momentum (SEM) tensor and $\kappa$ the coupling constant. Eq.(\ref{pfemt}) is applicable for a  so-called {\it perfect fluid}, sometimes called the {\it isotropic} perfect fluid (corresponding to a fluid without viscosity and heat conduction), where $\rho$ and $p$ are the energy density and pressure, respectively, as measured by an observer moving with the fluid; and $u_{\mu}$ is the fluid's four-velocity. One further element of the description of a perfect fluid is the assumed relation, called an {\em equation of state} $f(p,\rho)=0$ (EoS) between $p$ and $\rho$. In stellar physics, the so-called {\it polytropic} EoS, $p\propto \rho^{\gamma}$ is relevant; in cosmology, the isothermal\footnote{In the literature, this EoS is sometimes called the {\it barotropic} EoS or the {\it linear} EoS. However, most dictionaries give the meaning of {\it barotropic} as the property that the pressure depends on the density only, and {\it linear} would include relationships like $p = p_{0} +w\rho$, so we believe that the phrase {\it isothermal EoS} is more appropriate, in analogy with isothermal processes of the classical ideal gas where $pV =$ constant  } EoS, (\ref{eos}). In that context, $w=0$ describes the matter-dominated (or ``pressureless dust") case, \mbox{$w=1/3$} the radiation-dominated case, \mbox{$w<-1/3$} dark energy, and \mbox{$w<-1$} phantom energy. These latter  concepts have been introduced into cosmology in the last two decades  \cite{de,phantom}, after the discovery of the acceleration of the expansion of the universe \cite{acceleration-hiZsst,acceleration-SCP}.

  To find solutions for the contents and structure of a static spherically symmetric  spacetime, one usually starts with the ansatz~\cite[Sect.23.2]{mtw}
\begin{equation}
ds^{2} = -B(r) dt^{2} + A(r) dr^{2} + r^{2} d\Omega^{2}   \label{Schw-ansatz}
\end{equation}
for the line element, where $d\Omega^{2}$ is the line element for the unit sphere. For a static spherically symmetric  spacetime filled with an isotropic perfect fluid (SSSPF solutions), one must use (\ref{pfemt}) together with the staticity of the perfect fluid source, 
\begin{equation}
u^{\mu} = u^{0} \delta_{0}^{\mu}  \Longrightarrow u^{0} = B(r)^{-1/2}. \label{staticity}
\end{equation}
Given the metric ansatz (\ref{Schw-ansatz}), the  functions $B(r)$ and $A(r)$ are two of the four functions comprising the solution, $\rho(r)$ and $p(r)$ the other two.

The usual formalism leading to a solution (especially a numerical one) is a path via the well-known Oppenheimer-Volkoff (OV) equation~\cite{ov} (see \cite{OvTovSemiz} for arguments on the naming), 
where an auxiliary function $F(r)$ proportional to the "mass function" is introduced; and using the EoS and the differential relation between $F(r)$ and $\rho(r)$, one gets a differential equation for $F(r)$. If that equation can be solved, the metric functions can be found via relations between $A(r)$, $B(r)$ and $F(r)$.

However, for most EoS it is very hard to find an analytical solution, and this includes the deceptively simple isothermal (proportional) EoS $p=w\rho$, except for the trivial cases of $w=-1$ and $w=0$; and the case $w=-1/3$ \cite{CSB}. A solution for general $w$ would describe the spacetime around a spherically symmetric object embedded in dark energy or quintessence\footnote{Some authors use {\it quintessence} as a synonym for  dark energy, while others limit the use of that expression to the range $-1/3>w>-1$.} (note again that the Kiselev spacetime~\cite{Kiselev} does not qualify~\cite{Visser_on_Kiselev,Lake_on_Kiselev,Semiz_on_Kiselev}); but even if found (after all, a solution can always be found numerically), it could not describe a blackhole (see \cite{revisited,Semiz_on_Kiselev}) since the staticity condition $u^{\mu} = u^{0} \delta_{0}^{\mu}$ is not valid  inside the horizon.

For this EoS, the OV prescription gives the nonlinear equation
\begin{equation}
(w+1) F' (wrF'+F) + 2 w (rF''-2F')(r-F) = 0                      \label{Feq}
\end{equation}
where we put no constraint on $w$ other than that it is a constant. We repeat that its  solution for arbitrary $w$ is not known, in fact, not even for some given value of $w$. The triviality of the cases $w=-1$ and $w=0$ mentioned above can be clearly seen here, since for any of these values, half of eq.(\ref{Feq}) vanishes and the other half factorizes into simple linear equations. In \cite{allpoly}, we list {\it all} solutions where $F(r)$ is a finite polynomial, but none of the solutions is general, that is, they do not contain the number of arbitrary constants they should.

This problem, $w={\rm const}$, is treated e.g. in~\cite{Ivanov,Liouvillian}. In~\cite{Ivanov}, Ivanov concludes that the problem leads to the Abel equation of second kind, and hence the integrable cases are $w=0,-1,-1/3,-1/5$ and a particular solution for $w$ outside the interval $(-3-\sqrt{2},-3+\sqrt{2})$. The author also points out that the $w=-1$ and $w=-1/5$ cases are ``reciprocals'' of each other under the Buchdahl transformation~\cite{Buchdahl} (the other cases mentioned are self-reciprocal) and derives a $w=-1/5$ solution by Buchdahl-transforming the de Sitter solution. However, since he does not transform the Schwarzschild-de Sitter solution (the most general $w=-1$ solution), he does not find the most general $w=-1/5$ solution. 

Apparently, this is due to the author taking the mass function as a definite integral, his eq.(10). However, it is possible to motivate the definition of that function by realizing that the expression for $\rho$ coming from the 0-0 component of the Einstein Equations resulting from the line element (\ref{Schw-ansatz}) becomes a total derivative after multiplication by $r^2$; hence to use the indefinite integral. As result of this choice, the author also misses the Schwarzschild solution for $w=0$ (his $n=\infty$), writing ``what remains is trivial flat space–time'', and the black hole-like solution in the Einstein static universe, found in \cite{CSB}. In other words, the author misses the integration constant, which represents a mass point at the origin.

In \cite{Liouvillian}, the authors convert the same problem into a Lotka-Volterra differential system and characterize the Liouvillian integrability of this system using Darboux theory. They confirm integrability of the $w=0$, $w=-1$, $w=-1/3$ and $w=-1/5$ cases, and claim that the case $w=-1/6$ is also integrable. They define the mass function by the same definite integral as \cite{Ivanov}, and therefore mention the de Sitter and Einstein static universe as the $w=-1$ and $w=-1/3$ solutions, respectively; however it is not clear to the present author if their complicated formalism really misses the mass points at the origin despite the quoted definite integral, especially since they do not display any explicit line elements.

In this work, we solve the $w=-1/5$ SSSPF case completely, and show Buchdahl transform connections to the $w=-1$ solutions, and identify the Ivanov solution as a special case of the general one we find here.


\section{Solution for $w=-1/5$}
\label{Sol}

While the OV-formalism is well-known, the form (\ref{Schw-ansatz}) for the line element is not unique: it implies that the radial coordinate $r$ has been defined as $1/2\pi$ times the circumference of the largest circle lying in a given surface of spherical symmetry. The radius of that circle could be defined as any function (preferably monotonic) $y(r)$ of $r$, giving rise to the most general diagonal SSS line element
\begin{equation}
ds^{2} = -B(r) dt^{2} + A(r) dr^{2} + y(r)^{2} d\Omega^{2}.   \label{gen-ansatz}
\end{equation}

  Now, a {\it coordinate condition} can be chosen to define a valid coordinate system. For example, the coordinates associated with choice $y(r)=r$, as in eq.(\ref{Schw-ansatz}) are called Schwarzschild, or curvature coordinates. It turns out that our problem can be solved for $w=-1/5$ in {\it Buchdahl coordinates}\footnote{not particularly related to the Buchdahl transformation, although the transformation {\it does} keep the Buchdahl coordinate condition intact.} (see the table in \cite{Semiz_on_Kiselev} for examples of coordinate conditions) where $A(r)B(r) = 1$, as described below.
  
In these coordinates [eliminating $A(r)$], the EFE give  
\begin{subequations}
\begin{eqnarray}
\kappa \rho & = &   \frac{1}{y^{2}} - \frac{By'^{2}}{y^{2}} - \frac{y'B'}{y} - \frac{2By''}{y}  \label{ee00_B} \\ 
\kappa p & = &  -\frac{1}{y^{2}} + \frac{By'^{2}}{y^{2}} + \frac{y'B'}{y} \label{ee11_B} \\ 
\kappa p & = &  \frac{B'y'}{y} + \frac{B''}{2} + \frac{By''}{y},    \label{ee22_B}
\end{eqnarray}
\end{subequations}
for the 00, 11 and 22 components, respectively (the 33 component is the 22 component multiplied on both sides by $\sin^{2}\theta$). The prime of course denotes derivative with respect to $r$. Adding the first, second, and (four times) third equations after multiplication by $y/2$ gives
\begin{equation}
yB''+ 2y'B'+ y''B  = \kappa\frac{y}{2} (\rho + 5p),   \label{rho+5p_B}
\end{equation}
which obviously can easily be solved if $(\rho + 5p)=0$, giving
\begin{equation}
y(r)B(r) = C_{0} + C_{1} r,   \label{yB}
\end{equation}
where $C_{0}$ and $C_{1}$ are constants.   
  
Now we eliminate pressure from eqs.(\ref{ee11_B}) and (\ref{ee22_B}) to get what is usually referred to as the ``pressure isotropy equation'', then eliminate $B$ and its derivatives via (\ref{yB}) to get
\begin{equation}
(C_{0} + C_{1} r) y'' - 2 C_{1} y' + 2 = 0.   \label{y-DE}
\end{equation}
A relevant singular point of this differential equation is at $r_{0}=-C_{0}/C_{1}$, which becomes infinite in the case $C_{1}=0$, changing the nature of the equation. Therefore we need to consider the two cases separately.

\subsection{The generic case $C_{1} \neq 0$ }

In this case, eq.(\ref{y-DE}) can be solved to give
\begin{eqnarray}
y(r) &=& C_{2} +  \frac{r}{C_{1}} +  \frac{C_{3}(C_{0} + C_{1} r)^{3}}{3 C_{1}}      \nonumber \\
&=& \frac {3 (C_{1}C_{2}+r)+C_{3}(C_{0} + C_{1} r)^{3}} {3 C_{1}} \equiv 
\frac {f(r)} {3 C_{1}}  \label{y-GenSol}  
\end{eqnarray}
where $f(r)$ has been defined in the last step. This leads to the line element
\begin{equation}
ds^{2} = -\frac {3 C_{1} (C_{0} + C_{1} r)}{f(r)} dt^{2}  
+ \frac {f(r)} {3 C_{1} (C_{0} + C_{1} r)} \; dr^{2}  
+ \frac{f(r)^{2}}{9C_{1}^{2}} d\Omega^{2}.   \label{inSolGenMetric}
\end{equation}
However, some of the integration constants in this line element are redundant: We can transform $r \rightarrow C_1 \bar{r}-C_0/C_1$, $t \rightarrow \bar{t}/C_1$, redefine the constants $C_2 \rightarrow \bar{C_2} + C_0/C_{1}^{2}$, $C_3 \rightarrow 3\bar{C_3}/C_{1}^{5}$, then drop the overbars to get
\begin{equation}
ds^{2} = -\frac {r}{f(r)} dt^{2} + \frac {f(r)}{r} \; dr^{2}  + f(r)^{2} d\Omega^{2}   \label{finSolGenMetric}
\end{equation}
where $f(r)$ has been redefined as
\begin{equation}
f(r) = C_{2} + r + C_{3} r^{3}.   \label{fr}
\end{equation}
The density and pressure for this solution are given by 
\begin{eqnarray}
\rho(r)  &=& -\frac {15 \, C_{3} \, r^{2}}
{\kappa f(r)^{2}}   \label{GenSol-density} \\
p(r) &=& \frac {3 \, C_{3} \, r^{2}}{\kappa f(r)^{2}}.
   \label{GenSol-pressure}
\end{eqnarray}

\subsection{The particular case $C_{1}=0$}

In this case, eq.(\ref{y-DE}) easily gives
\begin{equation}
y(r) = C_{2}  + C_{3} r - \frac{r^{2}}{C_{0}} =   \frac {C_{0}(C_{2}+C_{3}r) - r^{2}} {C_{0}} \equiv \frac {g(r)} {C_{0}},  \label{y-PartSol}
\end{equation}
resulting in the line element
\begin{equation}
ds^{2} = -\frac {C_{0}^{2}}{g(r)} dt^{2} + \frac {g(r)} {C_{0}^{2}} dr^{2}  
+ \frac {g(r)^{2}} {C_{0}^{2}}  d\Omega^{2}.   \label{inSolPartMetric}
\end{equation}
Note that eqs.(\ref{y-PartSol}) and (\ref{inSolPartMetric}) do not follow as the $C_{1} \rightarrow 0$ limit of eqs.(\ref{y-GenSol}) and (\ref{inSolGenMetric}). Here also redundant constants can be removed. The appropriate transformation is $g(r) \rightarrow C_0 \bar{r}$, $t \rightarrow \bar{t}/\sqrt{C_0}$ and constant redefinition is $C=C_0C_{3}^{2}+4C_2$, and after dropping the overbars results in
\begin{equation}
ds^{2} = -\frac {dt^{2}}{r}  + \frac {r} {C-4 \, r} dr^{2} + r^{2}  d\Omega^{2}.   \label{finSolPartMetric}
\end{equation}
which is Solution 8a of \cite{allpoly} (which in turn is a special case of Tolman V \cite{Tolman} solution), after time rescaling. The density and pressure for this solution are given by 
\begin{eqnarray}
\rho(r) &=& \frac {5}{\kappa r^{2}}   \label{PartSol-density} \\
p(r) &=& -\frac {1}{\kappa r^{2}}. 
   \label{PartSol-pressure}
\end{eqnarray}
Interestingly, these expressions do not contain the constant $C$.

\subsection{Connections to the $w=-1$ case}

Let us first discuss the Ivanov's $w=-1/5$ solution, \cite{Ivanov}, eqs.(96) and (97), given in isotropic coordinates. To compare it with our solution, we must transform it into Buchdahl coordinates, which gives
\begin{equation}
ds^{2} = -\frac {dt^{2}}{1+4\,c\,r^2}  + (1+4\,c\,r^2) \, dr^{2} + (1+4\,c\,r^2)^2 \, r^{2}  d\Omega^{2}.   \label{IvanovBuchdahlC}
\end{equation}
It can be seen that this is the $C_2=0$ special case of our solution, (\ref{finSolGenMetric}), i.e. contains only one parameter. This is not surprising, since the solution was found in \cite{Ivanov} as the Buchdahl transform of the de Sitter solution, which also contains one parameter.

Could the general solution (\ref{finSolGenMetric}) have been found as the Buchdahl transform of the Schwarzschild-de Sitter --aka K{\"o}ttler-- solution (SdS/K)? While we initially derived (\ref{finSolGenMetric}) without utilizing the Buchdahl transform, we later found that the answer is affirmative. Starting from the well-known standard form $ds^2 = -F(r) dt^2 + F(r)^{-1} dr^2 + r^{2}  d\Omega^{2}$, with $F(r) = 1 - \frac{2M}{r} - \frac{\Lambda}{3} r^2$ (for this solution, Schwarzschild and Buchdahl coordinates coincide), the Buchdahl transform gives~\cite{Buchdahl}
\begin{eqnarray}
ds^{2} &=& -F(r)^{-1} dt^{2} + F(r)^{2} \left( F(r)^{-1} dr^2 + r^{2}  d\Omega^{2} \right)   \nonumber \\
 &=& -\frac{r}{r-2M-\Lambda r^3/3} dt^{2} + \frac{r-2M-\Lambda r^3/3}{r} dr^2 + (r-2M-\Lambda r^3/3)^{2}  d\Omega^{2},  
\label{SdS_BuchdahlT}
\end{eqnarray}
which agrees with (\ref{finSolGenMetric}) after identifications $C_2 \rightarrow -2M$, $C_3 \rightarrow -\Lambda/3$. One possible reason for the author of \cite{Ivanov} missing this solution was identified near the end of Sect.~\ref{sect:Intro} as the author defining the mass funtion as a definite integral. Another reason could be the author's unnecessary insistence on using the Buchdahl transform strictly in isotropic coordinates. The line element of the Schwarzschild-de Sitter/K{\"o}ttler spacetime cannot be written in terms of elementary functions in that coordinate system (see \cite{Solanki} for a recent treatment), the author of \cite{Ivanov} may have thought that the corresponding $w=-1/5$ solution cannot be written in terms of elementary functions either.

Finally, what about the particular solution, (\ref{finSolPartMetric})? It is not an obvious limiting case of (\ref{finSolGenMetric}), therefore it seems that it cannot be the Buchdahl transform of a special or limiting case of the SdS/K solution. What is it the Buchdahl transform of, then? For the answer, we can Buchdahl-transform the solution, since the transform is its own inverse. We get
\begin{equation}
ds^{2} = -r dt^{2} + \frac {1} {r (C-4 \, r)} dr^2 + d\Omega^{2},   
\label{Part_BuchdahlT}
\end{equation}
one of infinitely many representations of the Nariai spacetime \cite{Nariai1,Nariai2,EdNoteNariai} with $\Lambda=1$. The Nariai spacetime with arbitrary $\Lambda$ can be described by
\begin{equation}
ds^{2} = -r dt^{2} + \frac {1} {\Lambda r (C-4 \, r)} dr^2 + \frac{1}{\Lambda} d\Omega^{2},   
\label{Nariai}
\end{equation}
and satisfies $w=-1$. However the constancy of the coefficient of $d\Omega^{2}$ prevents this line element from being transformed into the SdS/K form by a nonsingular transformation of the radial coordinate; hence it is not surprising that our particular solution is distinct from the generic one (It turns out that the Nariai solution is a very nonobvious limiting case of the SdS/K solution, reachable by a singular tranformation \cite{Nariai_limit}). The Buchdahl transform of the general Nariai solution (\ref{Nariai}) gives
\begin{equation}
ds^{2} = -\frac {dt^{2}}{r} dt^{2} + \frac {r} {\Lambda (C-4 \, r)} dr^2 + \frac{r^{2}}{\Lambda} d\Omega^{2},   
\label{NariaiBuchdahl}
\end{equation}
which can immediately be transformed into (\ref{finSolPartMetric}) by simple scaling of coordinates and redefinition of constants. Hence we would seem to have found a degeneracy of the Buchdahl transform.

\section{Discussion of the spacetimes}

One advantage of the Buchdahl coordinates is that they guarantee correct signature, since $g_{00}$ and $g_{11}$ switch sign at same $r$-values $r=r_s$, if at all. These values may represent spatial points [if $f(r_s)=0$] or surfaces where spacetime may or may not be singular. If $r=r_s$ is a nonsingular surface, particles/observers can be imagined to cross it. However, on one side of such a surface the $r$ coordinate is spacelike, hence the spacetime is static; on the other side $r$ is timelike, hence the spacetime is dynamic. Particles entering these dynamic regions by crossing the surface are unable to change the sign of their $\dot{r}$ (it can never cross zero since $u_\mu u^\mu=-1$) later, hence cannot return; so these surfaces constitute {\it horizons}. Therefore in  some cases, these spacetimes could represent blackholes, as far as particle motion is concerned; {\it  if the solution is taken as valid on both sides.}

However, in the dynamic regions, the staticity equation (\ref{staticity}) is obviously not valid any more, hence the metric expressions for these regions do not satisfy Einstein's Equations with a perfect fluid source\footnote{except for the cases of vacuum ($\rho =0, p=0$) and $\Lambda$-fluid ($p+\rho=0$).}. Elsewhere~\cite{revisited} we expressed this fact as ``No new `simple' blackholes''. Therefore we are interested in the static regions only (where $A(r)$ and $B(r)$ are positive, i.e. $f(r)$ and $r$ have same sign), to whose discussion we now turn for the two cases.

\subsection{The generic spacetimes}

\begin{figure}[h!]
\caption{Graphs of $f(r)$ for given $|C_3|$ but different $C_2$'s. The three blue (or, generally rising) curves have positive $C_3$ and the eight red (or, generally falling) curves negative $C_3$. In each family, the values of $C_2$ decreases from top to bottom, corresponding to the cases discussed below and shown in the tables.} 
\centering
\includegraphics[width=0.5 \columnwidth]{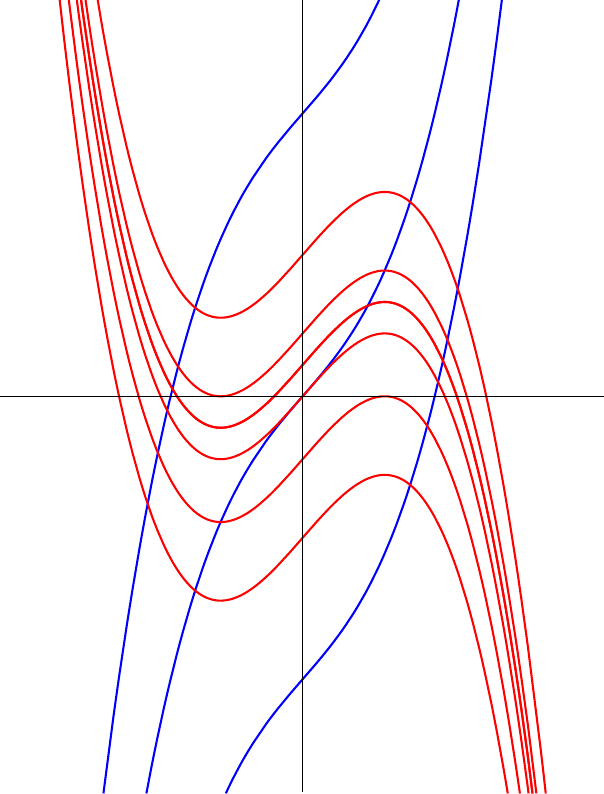}
\label{fig:figure} 
\end{figure}

The structure depends on the roots of $f(r)$ and $r$.
The scalar curvature vanishes at $r=0$, and diverges at the root(s) of $f(r)$; hence the roots of $f(r)$ represent singularities, not surprising since $f(r)$ represents the area radius of the surfaces of spherical symmetry.
Being third order, $f(r)$ can have one root which we will denote by $\pm r_*$ in the following; two roots to be denoted by $\pm r_a$ and $\pm r_b$, with an bar over the label $a$ or $b$ showing which one is the double root; or three roots to be denoted by $\pm r_1$, $\pm r_2$ and $\pm r_3$.
The sign of the root can be known (see Figure \ref{fig:figure}), and will be indicated explicitly, i.e. $r_*$, $r_a$, $r_b$, $r_1$, $r_2$ and $r_3$ will be positive quantities.
We need to distinguish cases based on the signs of $C_3$ and $C_2$; however, we should note that simultaneously changing the signs of $C_2$ and $r$ gives back the same solution. 

\subsubsection{Positive $C_3$}

See Table \ref{tab:positiveC3}. In this case, $f(r)$ has only one root $\pm r_*$, since its derivative never vanishes. The root has sign opposite of $C_2$ (see Figure \ref{fig:figure}).
\begin{table}[h!]
\caption{Spacetimes for positive $C_3$. In these cases, $f(r)$ has exactly one root, since  $f'(r)$ is strictly positive.  The sign of the root is explicitly indicated, i.e. the quantity $r_*$ is positive. $\;\,$ trf: The transformation $r \longrightarrow -r$ together with $C_2 \longrightarrow -C_2$,  H: horizon, $A$: area, NS: pointlike naked singularity.} 
\centering
\begin{tabular}{|p{0.16\linewidth}|p{0.13\linewidth}|p{0.17\linewidth}|p{0.11\linewidth}|p{0.21\linewidth}|p{0.14\linewidth}|}
\hline
\textbf{Sign of $C_2$; sign of root of $f(r)$} & \textbf{Region (possible coord. trf.)} & \textbf{Inner boundary} & \textbf{Outer boundary} & \textbf{Comment}  & \textbf{Suggested name} \\ \hline

\multirow{2}{*}{\parbox{\linewidth}{$\,$ \linebreak $C_2$ is positive; root is negative ($-r_{*}$)}} & Right & H at \linebreak $ r=0$, with $A=4\pi C_{2}^{2}$ & \centering $+\infty$ & infinite &  H$\infty$ \\ \cline{2-6} 

 & Left (trf) & NS at $-r_*$ $\stackrel{\rm trf}  {\longrightarrow}$ \linebreak NS at $+r_*$ & $-\infty$ $\stackrel{\rm trf}{\longrightarrow}$ $+\infty$ & infinite & S$\infty$   \\ \hline 
  
 \multirow{2}{*}{\parbox{\linewidth}{$C_2 = 0$  $\Longrightarrow$  root = 0 }} & Right  & $ r=0$, \linebreak regular, \linebreak $A=0$  & $+\infty$ & \multirow{2}{*}{\parbox{\linewidth}{One may consider the two $\infty$ regions to be connected by an infinitesimal wormhole, see text}} & \multirow{2}{*}{\parbox{\linewidth}{Two copies of 0$\infty$; or $\infty 0 \infty$ if considered conected }}   \\ \cline{2-4}
 
 & Left (trf) & $ r=0$, regular, $A=0$  & $-\infty$ $\stackrel{\rm trf}{\longrightarrow}$ $+\infty$ &  &    \\ \hline 
 
\multirow{2}{*}{\parbox{\linewidth}{$C_2$ is negative; root is positive ($+r_{*}$)}} & Right & $+r_*$  & $+\infty$  & \multicolumn{2}{c|}{Same as S$\infty$}  \\ \cline{2-6} 
 
  &  Left (trf) & 0 & $-\infty$ $\stackrel{\rm trf}{\longrightarrow}$ $+\infty$ & \multicolumn{2}{c|}{Same as H$\infty$}  \\ \hline 
 
\end{tabular}
\label{tab:positiveC3}
\end{table}

\paragraph{Positive $C_2$.}

In this subcase, $f(r)$ and $r$ are both negative for $r<-r_*$ (``Left'') and both positive for $r>0$  (``Right'').

\noindent \underline{Right region.} This spacetime is bounded by a horizon-like surface at $r=0$ (with finite spherical area $4 \pi C_2^2$), extending to $r \rightarrow \infty$, is not asymptotically flat (as the proper radial coordinate $\bar{r} \rightarrow \infty$, the volume increment $dV$ corresponding to $d\bar{r}$ is $32 \pi \sqrt{C_3} \bar{r}^3 d\bar{r}$ instead of $4 \pi \bar{r}^2 d\bar{r}$); and may be called the H$\infty$ spacetime. \\
\noindent \underline{Left region.} This spacetime is bounded by a naked singularity at $r=-r_*$ (with zero spherical area, i.e. a point), extending to $r \rightarrow -\infty$, not asymptotically flat (asymptotic properties similar to the right region). In fact, the above-mentioned $r \rightarrow -r$ transformation  will bring the solution into a form with negative $C_2$ (positive $C_3$) and $r>r_*$. It may be called the S$\infty$ spacetime.

\paragraph{Vanishing $C_2$.}

In this subcase, the root of $f(r)$, where the scalar curvature is expected to diverge, coincides with the root of $r$, where the scalar curvature is expected to vanish, giving finite scalar curvature. The left and right regions can be said to touch at their respective centers of symmetry without any singularity, which effectively constitutes a wormhole between the two regions, albeit with vanishingly small throat radius. Of course, this could also be said of the trivial Minkowski space, if $r$ in spherical coordinates is allowed to vary in $(-\infty,+\infty)$, but the case at hand is the limit of either of the two cases above, where two regions can be said to exist. We may call this case the $\infty 0 \infty$ spacetime.

\paragraph{Negative $C_2$.}

By the transformation mentioned above, this subcase gives spacetimes identical to the positive $C_2$ subcase (see Table \ref{tab:positiveC3}).

\subsubsection{Vanishing $C_3$}

In this case, $f(r)$ becomes $C_2+r$. The transformation $C_2+r \rightarrow r$ shows this spacetime to be the Schwarzschild spacetime with $C_2=2M$ (Positive $C_2$ gives the usual Schwarzschild blackhole, negative $C_2$ gives an asymptotically flat spacetime with a naked singularity at the center). 

\subsubsection{Negative $C_3$}

See Table \ref{tab:negativeC3}. In this case, $f(r)$ has one, two or three roots depending on if $|C_2|$ is larger than, equal to or smaller than $\frac{2}{3}\frac{1}{\sqrt{3|C_3|}} = C^{*}$  (see Figure \ref{fig:figure}).

\paragraph{Positive $C_2>C^{*}$.}

In this subcase, $f(r)$ has a single positive root $r_*$, so $A(r)$ and $B(r)$ are positive in the finite range $0<r<r_*$, where $r_* > \frac{1}{\sqrt{-3C_3}}$. One boundary, $r=0$, is again a horizon-like surface (finite spherical area $4 \pi C_2^2$), the other an origin-like naked sigularity. The transformation $r \rightarrow r_*-r$ will put the singularity at the origin and the horizon at $r=r_*$, so this spacetime consists of a naked singularity surrounded by a horizon; i.e. it is compact. Interestingly, the area radius increases as the radial distance from the point singularity increases, reaching a maximum $C_2+C^*$, and then {\it decreases} to $C_2$ at the horizon. This makes the topology of this spacetime the 3-D analog of a regular sphere with a circumpolar region (smaller than a hemisphere) removed. We may call this case the ${\rm SH}_{\rm L}$ spacetime since it is larger than the ${\rm SH}_{\rm L}$ spacetime introduced next. 

\paragraph{Positive $C_2=C^{*}$.}

In this subcase, $f(r)$ and $r$ are again positive in a finite range $0<r<r_b$ (see Figure \ref{fig:figure} and Table \ref{tab:negativeC3}) where $r_b=3C_*$. $r=0$ is horizon-like, $r=r_b$ is singular, the area radius has a maximum (equal to $2 C_2= 2 C^*$); so this spacetime is qualitatively similar to the last discussed spacetime; we may call it the ${\rm SH}_{\rm 1}$ spacetime, since it really has only one parameter, namely $C_3$.

\begin{landscape}
\begin{table}[]
\caption{Spacetimes for negative $C_3$. In these cases, $f(r)$ can have one, two or three roots. The sign(s) of the root(s) is/are explicitly indicated, i.e. the quantities $r_*$, $r_a$, $r_1$ etc. are  positive. Define $C^* = 2/\sqrt{-27 C_3}$. H: horizon, $A$: area, NS: pointlike naked singularity.} 
\centering
\begin{tabular}{|p{0.16\linewidth}|p{0.10\linewidth}|p{0.17\linewidth}|p{0.16\linewidth}|p{0.20\linewidth}|p{0.10\linewidth}|}
\hline
 \textbf{Sign of $C_2$; roots of $f(r)$ \& their signs}  & \textbf{Region/trf} & \textbf{Inner boundary} & \textbf{Outer boundary} & \textbf{Comment}  & \textbf{Suggested name} \\ \hline

$C_2>C^*$;   $r_*>3 C^*$  & trf: $r \longrightarrow r_* - r$ & H at $r=0$, with $A=4\pi C_{2}^{2}$, $\stackrel{\rm trf}{\longrightarrow}$  NS at $r=0$. & NS  at $r=r_*$,  $\stackrel{\rm trf}{\longrightarrow}$ H with $A=4\pi C_{2}^{2}$ at $r=r_*$.  & Compact, topology like 3-D sphere with a circumpolar region removed  & ${\rm SH}_{\rm L}$ (since larger than ${\rm SH}_1$)  \\ \hline

$C_2=C^*$; $-r_{\bar a}=-\frac{3}{2}C^*$ (double), $r_{b}=3C^*$ & trf: $r \longrightarrow r_b - r$ & H at $ r=0$, with $A=4\pi C_{2}^{2}=4\pi C_{*}^{2}$, $\stackrel{\rm trf}{\longrightarrow}$ NS at $ r=0$.  & NS  at $ r=r_b$,  $\stackrel{\rm trf}{\longrightarrow}$ H with $A=4\pi C_{2}^{2}=4\pi C_{*}^{2}$ at $ r=r_b$.  & Compact, one-parameter, similar topology to ${\rm SH}_{\rm L}$, smaller for given $C_3$ & ${\rm SH}_1$ (since one-parameter)  \\ \hline
 
\multirow{2}{*}{\parbox{\linewidth}{$\,$ \linebreak $0<C_2<C^*$; $-r_1$, $-r_2$,  $r_3$}}   & Right (trf: $r \longrightarrow r_3 - r$) & H at $ r=0$, with $A=4\pi C_{2}^{2}$ $\stackrel{\rm trf}{\longrightarrow}$  NS at $ r=0$. & NS at $r_3$ $\stackrel{\rm trf}{\longrightarrow}$ H with $A=4\pi C_{2}^{2}$ at $ r=r_3$.  & Compact, similar topology to ${\rm SH}_1$, but smaller & ${\rm SH}_{\rm S}$ (smaller than ${\rm SH}_1$)  \\ \cline{2-6} 
 
 & Left (trf: $r \longrightarrow -r$) &  NS at $-r_1$, $\stackrel{\rm trf}{\longrightarrow}$ NS at $+r_1$  &  NS at $-r_2$, $\stackrel{\rm trf}{\longrightarrow}$ NS at $+r_2$    & Compact, closed & SS  \\ \hline
 
\multirow{2}{*}{\parbox{\linewidth}{$C_2=0$; $-r_1=\frac{-1}{\sqrt{-C_3}}=-\frac{3\sqrt{3}}{2}C^{*}$, $r_2=0$,  $r_3=\frac{1}{\sqrt{-C_3}}=\frac{3\sqrt{3}}{2}C^{*}$}}  & Right \linebreak  & $ r=0$, regular, $A=0$  &  NS at $r_3$  & \multirow{2}{*}{\parbox{\linewidth}{One may consider the two compact regions to be connected by an infinitesimal wormhole, see text}}  & \multirow{2}{*}{\parbox{\linewidth}{Two copies of 0S; or S0S if considered connected}}  \\ \cline{2-4}

  & Left (trf: $r \longrightarrow -r$)  & $ r=0$, regular, $A=0$ & NS at $-r_3$,  $\stackrel{\rm trf}{\longrightarrow}$ NS at $+r_3$ &   &   \\ \hline
  
\multirow{2}{*}{\parbox{\linewidth}{$-C^*<C_2<0$, $-r_3$, $r_2$, $r_1$ }} & Right   & \multicolumn{4}{c|}{Same as SS, without the need of a transform}   \\ \cline{2-6} 

  & Left & \multicolumn{4}{c|}{Same as ${\rm SH}_{\rm S}$ by $r \longrightarrow -r$}  \\ \hline
  
$C_2=-C^*$; $-r_{b}=-3C^*$, $r_{\bar a}=\frac{3}{2}C^*$ (double) & trf: $r \longrightarrow r - r_b$ & \multicolumn{4}{c|}{Same as ${\rm SH}_{\rm 1}$}  \\ \hline

$C_2 < -C^*$; $-r_{*} < -3C^*$ & trf: $r \longrightarrow r - r_*$ & \multicolumn{4}{c|}{Same as ${\rm SH}_{\rm L}$}  \\ \hline

\end{tabular}
\label{tab:negativeC3}
\end{table}
\end{landscape}

\paragraph{Positive $C_2<C^{*}$.}

In this subcase, $f(r)$ and $r$ have same sign in the two finite ranges $-r_1<r<-r_2$ (``Left'') and $0<r<r_3$ (``Right'').

\noindent \underline{Right region.} This spacetime is also qualitatively similar to the SH spacetimes discussed above, but is smaller, so we may call it the ${\rm SH}_{\rm S}$ spacetime.  \\
\noindent \underline{Left region.} The ends of the range $-r_1$, $-r_2$ are both pointlike naked singularities, with a maximum of $f(r)^2$ in the middle; hence the spacetime is a compact closed spacetime, topologically similar to a 3-sphere, with naked singularities at antipodal points, and can be called the SS spacetime.

\paragraph{Vanishing $C_2$.}

In this subcase, the roots of $f(r)$ become $-r_1=\frac{-1}{\sqrt{-C_3}}$, $r_2=0$ and $r_3=\frac{1}{\sqrt{-C_3}}$. All of these roots represent points (not surfaces), and $r=-r_1$ and $r=r_3$ are singularities, but $r=0$ is nonsingular, as in the above $\infty 0 \infty$ spacetime. Like in that case, we have two spacetimes touching at their respective centers of symmetry without any singularity, constituting an infinitesimal wormhole; unlike that case, the two spacetimes are compact, with naked singularities at the antipodes --or alternative centers--, so we may call it the S0S spacetime.

\subsection{The particular spacetime}

The form (\ref{inSolPartMetric}) for the line element of the particular spacetime has the advantages of Buchdahl coordinates mentioned in the beginning of this section, but multiple parameters; whereas the form (\ref{finSolPartMetric}) (which is not novel) lacks those advantages despite being simpler in the sense of having only one parameter. However, it is possible to have both: Defining $r \longrightarrow \bar{r} + C_0 C_3/2$, we get $g(r) \longrightarrow C_0 C_2 + C_0^2 C_3^2/4 - \bar{r}^2$. Since we need positive $g(r)$ for the spacetime to be static, $C_0 C_2 + C_0^2 C_3^2/4$ must be positive; and $\bar{r}$ must be between the two roots, a finite range. The  form (\ref{finSolPartMetric}) also shows a finite range, $0 < r < C/4$, but the ranges of the two forms do not match, since the transformation $g(r) \rightarrow C_0 \bar{r}$ from (\ref{inSolPartMetric}) to (\ref{finSolPartMetric}) is not one-to-one.

Further defining $C_0 C_2 + C_0^2 C_3^2/4 \longrightarrow \bar{C}^2$, then $\bar{r} \longrightarrow \bar{C} \tilde{r}$, and also rescaling time, we get
\begin{equation}
ds^{2} = \frac{\bar{C}^4}{C_0^2} \left[ -\frac {1}{1-\tilde{r}^2} d\tilde{t}^{2} + (1-\tilde{r}^2) dr^{2}  + (1-\tilde{r}^2)^2  d\Omega^{2}  \right]
 \label{PartMetricRationalized}
\end{equation}
showing a Buchdahl-form line element with a single (global scaling) parameter.

The form (\ref{PartMetricRationalized}) describes an $S^3$-like spatial topology, with two naked singularities at antipodal positions $\tilde{r} = \pm 1$, and better suited for interpretation than (\ref{finSolPartMetric}). The form (\ref{finSolPartMetric}) covers this spacetime twice: As we radially go from one root/singularity to the other in (\ref{PartMetricRationalized}), in (\ref{finSolPartMetric}) we must go from $0$ to $C/4$ {\it and back}; $r=C/4$ in (\ref{finSolPartMetric}) corresponds to the 3-D equator $\tilde{r} = 0$ of the $S^3$-like spatial part; it is not easy to see from the form (\ref{finSolPartMetric}) that $r=C/4$ is {\it not} singular.  This may not be too surprising since in the transformation to (\ref{finSolPartMetric}) we lose the Buchdahl form.  We can see that this spacetime is also compact, with finite proper extension in the radial direction and a finite proper volume. We have kept the discussion of the particular spacetime brief, and not included it in the tables because it is not novel, at least in the form (\ref{finSolPartMetric}).

\subsection{The conformal (Carter-Penrose) diagrams}

\begin{figure}[h!]
\caption{A Penrose diagram of the H$\infty$ spacetime.} 
\centering
\includegraphics[width=0.75 \columnwidth]{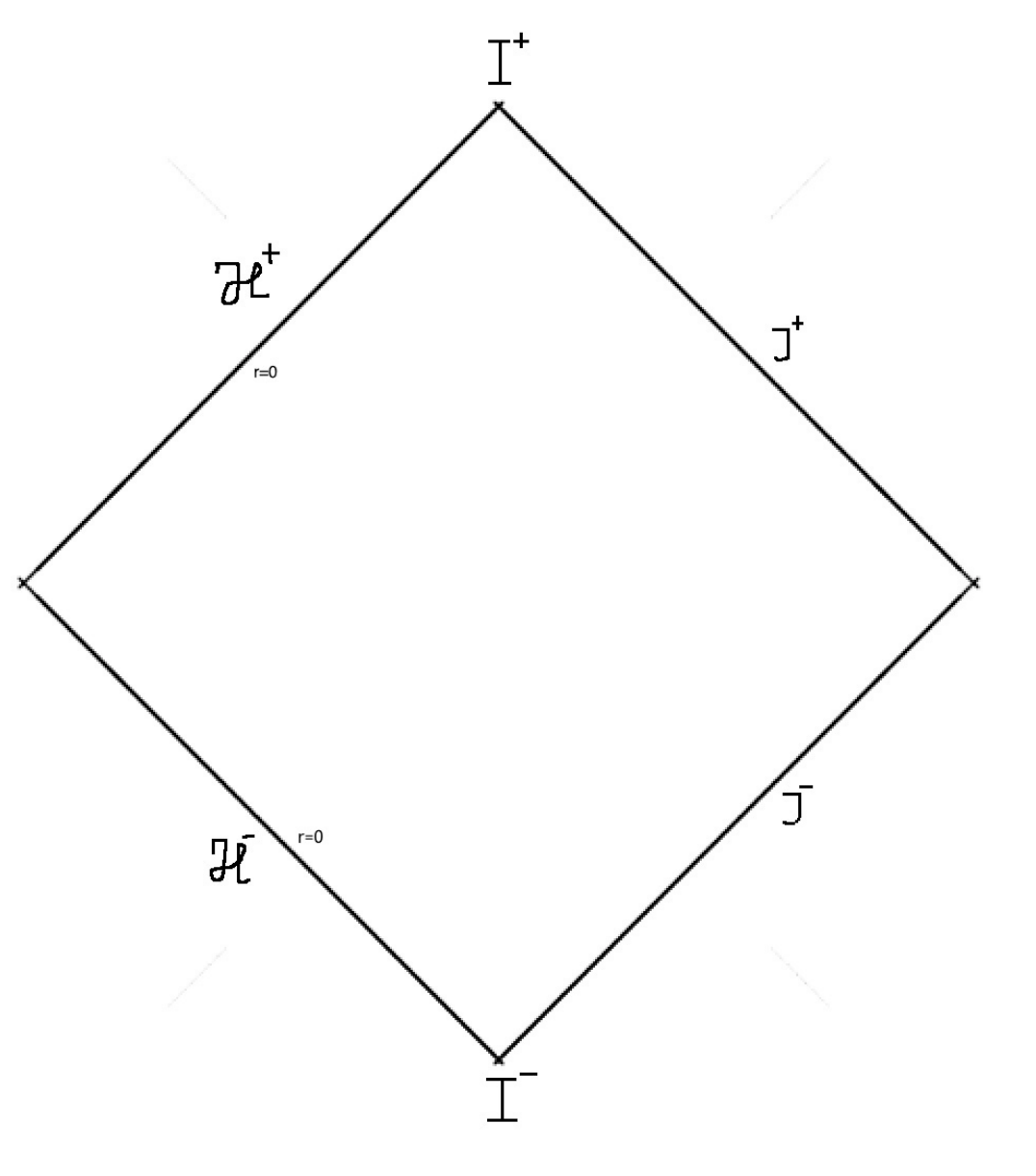}
\label{fig:horizons-1} 
\end{figure}

Suppressing the angular coordinates, the line element for the generic spacetime, (\ref{finSolGenMetric}), can be written as
\begin{equation}
ds^{2} = \frac {r}{f(r)} \left(-dt^{2} + d\bar{R}^{2}  \right)   \label{GenMetricConf1}
\end{equation}
where
\begin{equation}
\bar{R} = \int \frac {f(r)}{r} dr =  C_2 \ln r + r + \frac {C_3}{3} r^3, \label{ConfR}
\end{equation}
and $r$ and $f(r)$ in (\ref{GenMetricConf1}) are now considered implicit functions of $\bar{R}$ via (\ref{ConfR}). Now we can use the form (\ref{GenMetricConf1}) to construct conformal diagrams showing causal relationships. In particular, $r=0$ is mapped to $\bar{R} \longrightarrow -\infty$ for $C_2 > 0$ (or is made so by a transformation). These are the spacetimes with horizons, i.e. the H$\infty$, ${\rm SH}_{\rm L}$, ${\rm SH}_{\rm 1}$ and ${\rm SH}_{\rm S}$ spacetimes. Switching to null coordinates, transforming to null coordinates with finite range, discarding the conformal factor, transforming back to spacelike and timelike coordinates, we find that the horizon in each case appears in the diagram like the horizon in the right half of the Penrose diagram of the Schwarzschild spacetime; except that it is labeled $r=0$ instead of $r=2M$. For H$\infty$, the infinities exist, and are obviously similar to the Schwarzschild case for this purpose (even though this spacetime is not asymptotically flat), so the conformal (Carter-Penrose) diagram of H$\infty$ is identical in appearance to the right diamond of the corresponding diagram for the Schwarzschild spacetime (Figure \ref{fig:horizons-1}). 

\begin{figure}[h!]
\caption{A Penrose diagram of either the ${\rm SH}_{\rm L}$, or the ${\rm SH}_{\rm 1}$ or the ${\rm SH}_{\rm S}$ spacetime, and its transformed (flipped) version. The radial coordinate of the singularity becomes the radial coordinate of the horizon after the transformation.} 
\centering
\begin{picture}(380,250)
\put(0,0){\includegraphics{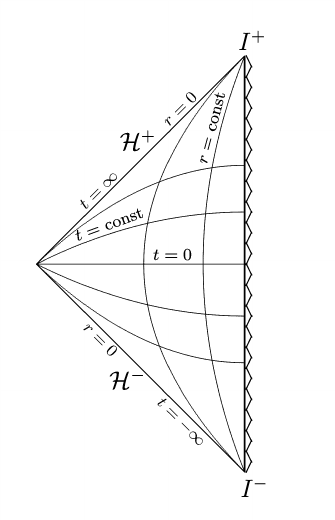}}
\put(250,0){\includegraphics{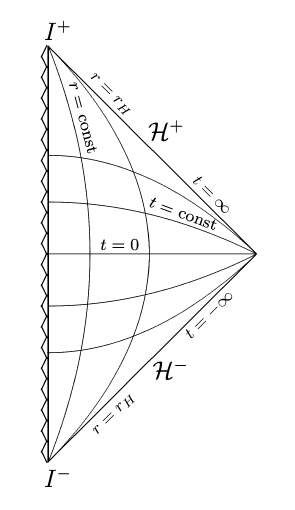}}
\put(125,110){\rotatebox{90}{\footnotesize $r=r_S$}}
\put(260,130){\rotatebox{-90}{\footnotesize $r=0$}}
\thicklines \put(140,125){\vector(1,0){105}}  \thinlines
\end{picture}
\label{fig:horizons-2} 
\end{figure}

The ${\rm SH}_{\rm L}$, ${\rm SH}_{\rm 1}$ and ${\rm SH}_{\rm S}$ spacetimes, however, do not extend to infinity; they end at a naked singularity each, at given $r$-values; so their Penrose diagrams will be cut off at some $r=$const. curve, which can be arranged to be vertical by the choice of unit for $r$. The transformations in Table \ref{tab:negativeC3} put the singularity at the origin, hence flip the Penrose diagram horizontally (Figure \ref{fig:horizons-2}).

\begin{figure}[h!]
\caption{A Penrose diagram of the S$\infty$ spacetime.} 
\centering
\includegraphics[width=0.4 \columnwidth]{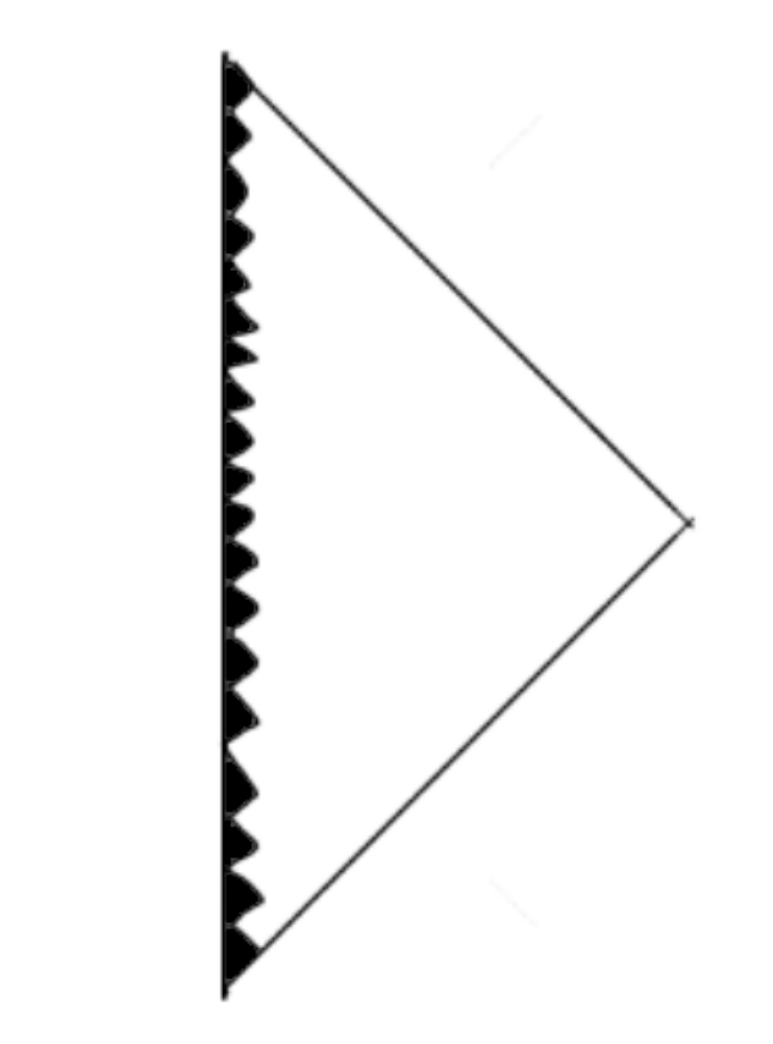}
\label{fig:horizons-3} 
\end{figure}

The Penrose diagram of the S$\infty$ spacetime is also a cut-off version of that of the H$\infty$ spacetime, this time the right half. It does have infinities, but no horizon; it has a naked singularity instead (Figure \ref{fig:horizons-3}). 

The Penrose diagram of the $\infty 0 \infty$ spacetime is identical in appearance to two Minkowski-space Penrose diagrams glued together (Figure \ref{fig:wormh}). The dotted line in the middle represents the infinitesimal wormhole at the origin, and should be thought of like a semi-permeable membrane: An object can either reflect off it, staying in its 0$\infty$  spacetime, or pass through it into the other. Or one may consider the probability of passing through the wormhole to be also infinitesimal, hence impose reflection, giving two unconnected spacetimes. 
\begin{figure}[h!]
\caption{A Penrose diagram of the $\infty 0 \infty$ spacetime. The dotted line in the middle will usually (always?) reflect objects.} 
\centering
\begin{picture}(240,240)
\put(20,115){\line(1,1){100}}	\put(20,115){\line(1,-1){100}}
\put(220,115){\line(-1,-1){100}}\put(220,115){\line(-1,1){100}}
\put(118,3){$I^{-}$}			\put(117,218){$I^{+}$}
\put(170,55){${\cal I}^{-}$}  	\put(170,170){${\cal I}^{+}$}
\put(145,193){\rotatebox{-45}{\scriptsize $r=\infty$}}	\put(180,158){\rotatebox{-45}{\scriptsize $t=\infty$}}
\put(145,34){\rotatebox{45}{\scriptsize $r=\infty$}}	\put(180,69){\rotatebox{45}{\scriptsize $t=-\infty$}}
\put(55,55){${\cal I}^{-}$}  	\put(60,170){${\cal I}^{+}$}
\put(80,182){\rotatebox{45}{\scriptsize $r=\pm \infty$}}	\put(40,141){\rotatebox{45}{\scriptsize $t=\infty$}}
\put(34,91){\rotatebox{-45}{\scriptsize $r=\pm \infty$}}	\put(77,48){\rotatebox{-45}{\scriptsize $t=-\infty$}}
\put(7,111){$I^{0}$}	\put(224,111){$I^{0}$}
\qbezier(20,115)(120,210)(220,115)	\qbezier(20,115)(120,20)(220,115)
\qbezier(20,115)(120,165)(220,115)	\qbezier(20,115)(120,65)(220,115)
\qbezier(20,115)(120,115)(220,115)
\put(38,126){\rotatebox{20}{\scriptsize $t=$ const}}
\put(75,117){\scriptsize $t=0$}
\qbezier(120,15)(23,115)(120,215)	\qbezier(120,15)(212,115)(120,215)
\qbezier(120,15)(80,115)(120,215)	\qbezier(120,15)(160,115)(120,215)
\put(130,196){\rotatebox{-75}{\scriptsize $r=$ const}}
\qbezier[70](120,15)(120,115)(120,215)
\put(122,138){\rotatebox{-90}{\scriptsize $r=0$}}
\end{picture}
\label{fig:wormh} 
\end{figure}

\begin{figure}[h!]
\caption{A conformal diagram (left, extending indefinitely up and down) and a Penrose diagram (right) of the SS spacetime.} 
\centering
\includegraphics[width=0.5 \columnwidth]{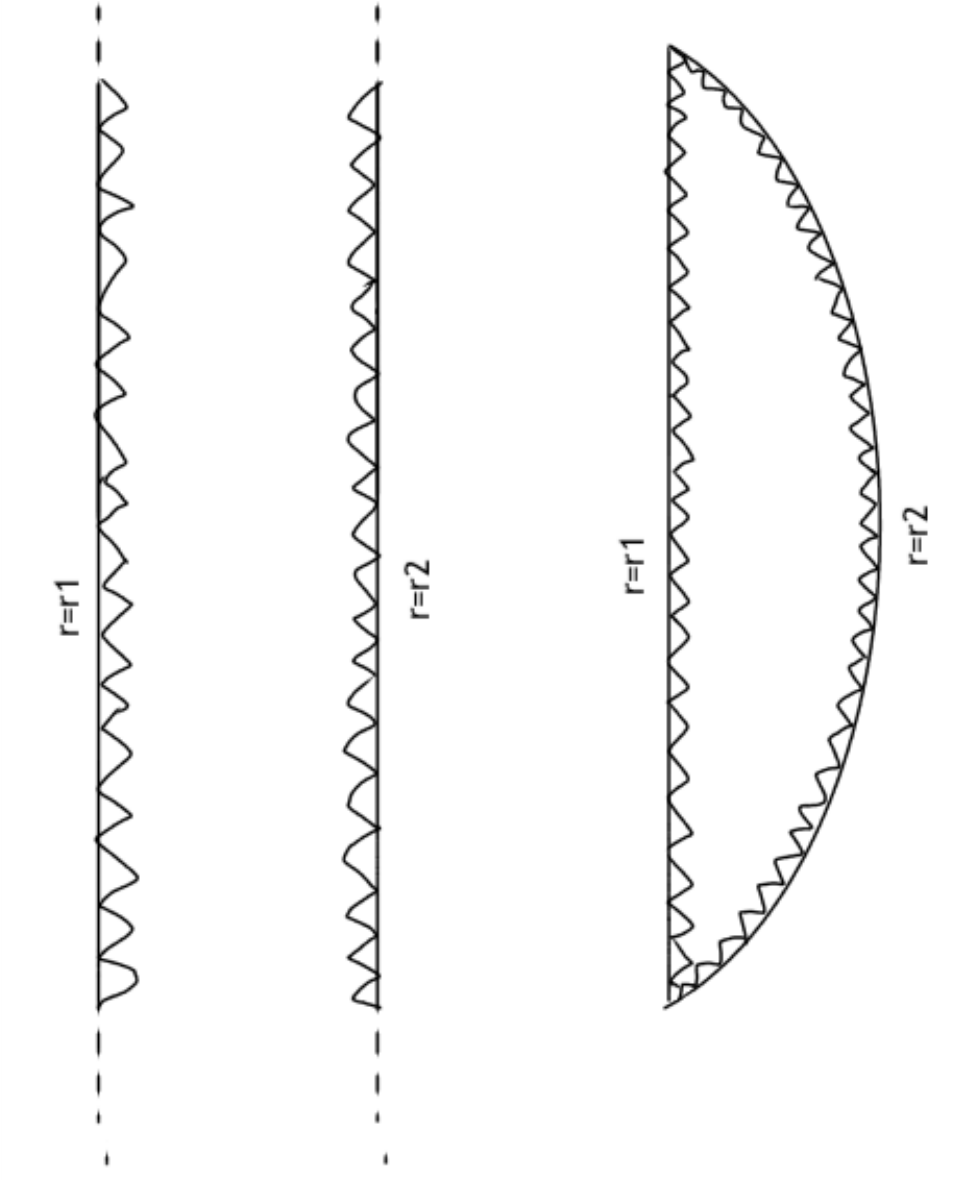}
\label{fig:ss} 
\end{figure}

For the SS spacetime, the variable $\bar{R}$ does not reach $\pm \infty$, even though $C_2 \neq 0$, since $r$ is confined between two negative (positive after transformation) values, hence does not reach zero or $\pm \infty$. Therefore, a photon or an object can cross the space between the two naked singularities in finite coordinate time, in contrast to the cases of Figures \ref{fig:horizons-1}-\ref{fig:wormh} (of course, the fact that a trip takes infinite coordinate time does not necessarily mean it takes infinite {\em proper} time, recall fall to horizon in Schwarzschild). Therefore, an infinite number of zigzags between singularities is possible for a photon, very obvious from a conformal diagram in the form of an infinite strip, but not so obvious from the Penrose diagram (Figure \ref{fig:ss}).

\begin{figure}[h!]
\caption{A conformal diagram (left) and a Penrose diagram (right), applying both to the S0S spacetime and to the particular spacetime. The dotted line represents an infinisimal wormhole for S0S, and the ``equator'' for the particular spacetime.} 
\centering
\includegraphics[width=0.5 \columnwidth]{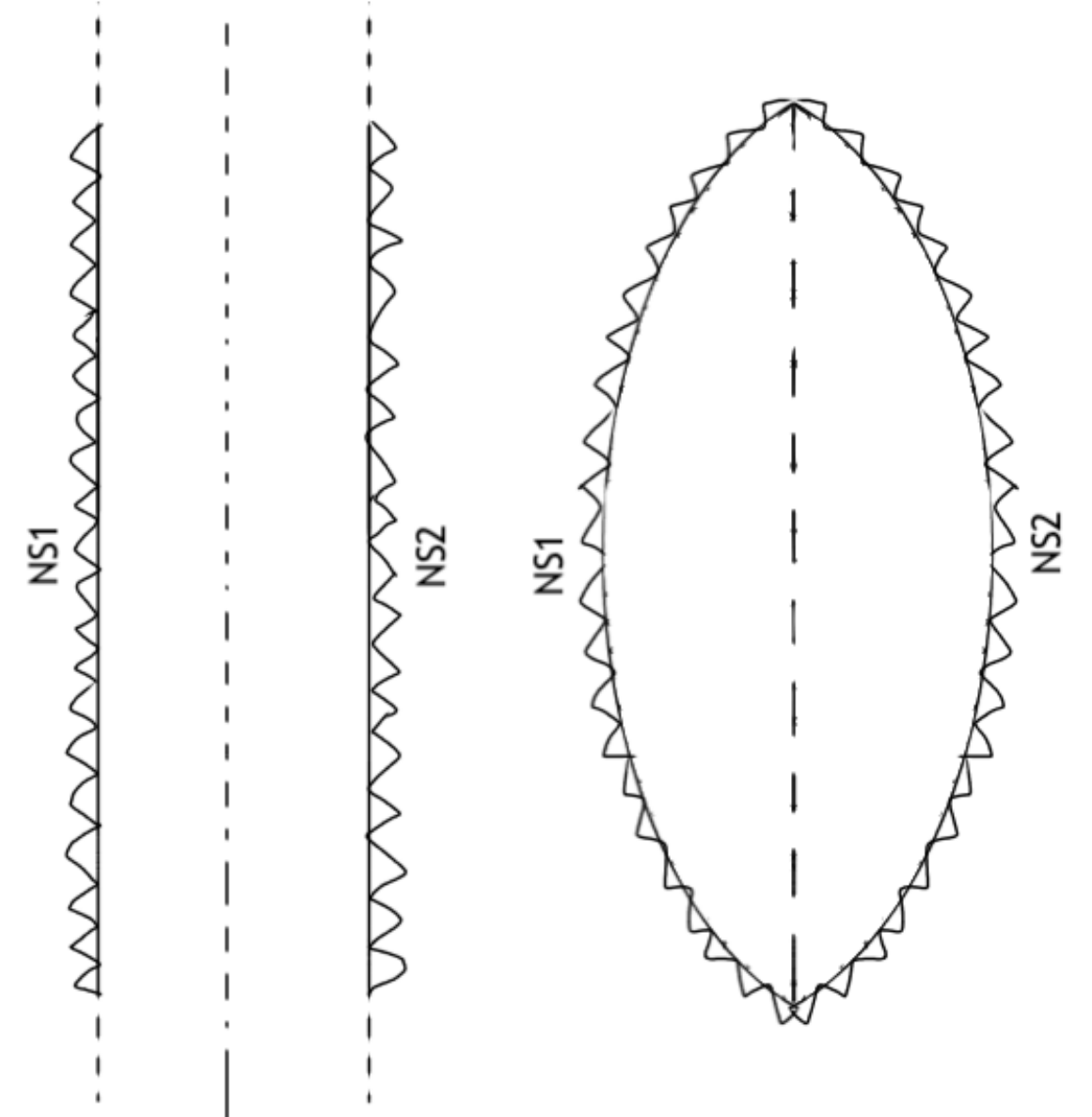}
\label{fig:s0s} 
\end{figure}

The S0S spacetime is like a truncated version of the $\infty 0 \infty$ spacetime. Again $\bar{R}$ does not reach $\pm \infty$, this time since $C_3 = 0$ and the space is truncated by the singularity; hence again infinite zigzagging is possible, and {\em two} diagrams. The origin has the same property as in the $\infty 0 \infty$ spacetime, hence the similar dotted line (Figure \ref{fig:s0s}).

Finally, the particular spacetime is also bounded by two naked singularities at certain values of the radial coordinate. Here also the variable for $\bar{R}$ (defined for making $ds^2$ proportional to $-dt^{2} + d\bar{R}^{2}$) changes by finite amount when traversing the space radially, so infinite zigzagging is possible. Therefore Figure \ref{fig:s0s} qualitatively also applies here, the dotted line for this spacetime representing ``the equator'' of the $S^3$-like topology.

\section{Summary and final comments}

We have presented the full solutions of Einstein's Equations for the static spherically symmetric case of a perfect fluid source with equation of state $p=-\rho/5$, exhibited connection to the $p=-\rho$ case, and discussed the resulting spacetimes for all ranges of the solutions' parameters. As discussed in the introduction, the motivation for the EoS $p = w \rho$ comes from cosmology, where $w$ values less than -1/3 are routinely used in discussions of accelerating expansion of the universe, and hence dark energy, even though the corresponding fluid is usually considered unphysical or exotic, since it violates various energy conditions. 

Despite the caveats associated with the nature of the negative-$w$ fluids, once they are seriously considered in one area, it is natural to wonder if applications --even if only at the level of thought experiments-- can be found in other areas, for example, if these fluids could play a role on e.g. galactic scales or contribute to or modify the structures of compact objects, "stars". As also discussed in \cite{Fazl-Fara}, the naked singularities at the origins of some of these spacetimes and the fact that these solutions cannot be matched to an external vacuum --Schwarzschild-- solution due to their nonvanishing pressure do not mean that the solutions are useless. In ``stellar'' solutions, one of the solutions could be used in a radial coordinate range that excludes the troublesome boundaries, and at the ends of that range, be matched to some other solutions, again, as also argued\footnote{This work, \cite{Fazl-Fara}, is an analysis of the spacetimes under discussion, based on a preliminary version of the present work. Unfortunately, the redundancies in the parameters of the line elements were not removed in that version [the forms (\ref{y-GenSol}) \& (\ref{inSolGenMetric}) and (\ref{y-PartSol}) \& (\ref{inSolPartMetric}) were reported], so the solutions seemed to have four parameters. Because of the apparent complexity of the solutions, the authors studied only some simple cases.} in \cite{Fazl-Fara}. In fact, it might be possible to match them at some surface with an exterior de Sitter or Schwarzschild - de Sitter solution since the negative pressure of the fluid could be made to match $-\Lambda$ at the surface, and this might be more realistic since our universe seems to have a cosmological constant or equivalent.

 The exotic properties of these fluids may be realized e.g. via various fields, or as effective fluids, as sometimes done in cosmology. There is a fair body of literature discussing the possibility of ``dark energy stars'' with EoS more extreme than the present case; and it is also possible that in a static situation, a "more physical" fluid and a possible cosmological constant together can be equivalent to a single negative-$w$ fluid, just like the Einstein static universe model where matter ($w=0$) and a cosmological constant $\Lambda$ ($w=-1$) combined in a certain ratio, are equivalent to a fluid with $w=-1/3$.

Very few exact SSSPF solutions are known for the EoS $p = w \rho$, and the present work may have completed all the fully exactly integrable $w$ cases (see end of Sect.~\ref{sect:Intro}). The spacetimes found here are either compact, bounded by horizons and/or naked singularities, or they do not have the correct asymptotic properties to be relevant in our universe\footnote{The asymptotic properties are neither Minkowski nor de Sitter; the infinite spacetimes also have negative energy densities.}. The metric for the case with positive $C_3$ and $C_2$ (the H$\infty$ spacetime) features a horizon, and could describe a blackhole in an infinite universe, but the dynamical region does not have a perfect fluid source, so that region is not valid in the present context.

However, the generic solutions with negative $C_3$ (SH types, SS and S0S; all of them compact) and the particular solution may not be as exotic as they first seem; they satisfy all the usually quoted energy conditions [dominant, strong, weak, null; see eqs. (\ref{GenSol-density}), (\ref{GenSol-pressure}), (\ref{PartSol-density}) and (\ref{PartSol-pressure})]. The generic solutions with positive $C_3$ (all of them infinite) violate all of them.

Finally, the EoS $p=-\rho/5$ by itself does not have a well-known interpretation, unlike the cases listed between eqs. (\ref{eos}) and (\ref{Schw-ansatz}), or $w=-1$ (equivalent to $\Lambda$) or $w=-1/3$ (equivalent to a gas of cosmic strings). Hence it is intriguing that the solution becomes simple/integrable in that case. Another final thought is that the surprisingly diverse set of spacetimes resulting from the solution of the present problem leads to the expectation of even more diverse set for the seemingly simple problem of static spherically symmetric configurations obeying the EoS (\ref{eos}), $p=w \rho$, and may dampen the expectation of a single reasonably elementary solution for that problem.

\section*{Acknowledgements}

I thank Fatih Akta\c{s} for useful discussions; and particularly would like to express my thanks to the anonymous referees of the journal for their constructive comments.

\end{document}